\documentclass{article}

\usepackage[final]{neurips_2022}

\usepackage[utf8]{inputenc} % allow utf-8 input
\usepackage[T1]{fontenc}    % use 8-bit T1 fonts
\usepackage{hyperref}       % hyperlinks
\usepackage{url}            % simple URL typesetting
\usepackage{booktabs}       % professional-quality tables
\usepackage{amsfonts}       % blackboard math symbols
\usepackage{nicefrac}       % compact symbols for 1/2, etc.
\usepackage{microtype}      % microtypography
\usepackage{xcolor}         % colors
\usepackage{multirow}
\usepackage{graphicx}
\usepackage{verbatim}
\usepackage{float}
\usepackage[group-separator={\,}]{siunitx}

\title{
What cleaves? Is proteasomal cleavage prediction reaching a ceiling?
}

\author{%
  Ingo Ziegler, $^1$
  Bolei Ma,$^1$
  Ercong Nie,$^1$ \\
  \textbf{Bernd Bischl,$^{2,3}$
  David Rügamer,$^2$
  Benjamin Schubert,$^{4,5}$
  Emilio Dorigatti$^{2,4}$} \\
  $^1$ Center for Information and Language Processing, LMU Munich, \\
  $^2$ Department of Statistics, LMU Munich, \\
  $^3$ Munich Center For Machine Learning, \\
  $^4$ Institute of Computational Biology, Helmholtz Zentrum München, \\
  $^5$ Department of Mathematics, TUM Munich \\
  \texttt{\{ziegler.ingo, bolei.ma\}@campus.lmu.de,} 
  \texttt{nie@cis.lmu.de}, \\
  \texttt{\{bernd.bischl, david.ruegamer, emilio.dorigatti\}@stat.uni-muenchen.de}\\
  \texttt{benjamin.schubert@helmholtz-muenchen.de}
}

\begin{document}

\maketitle

\begin{abstract}
Epitope vaccines are a promising direction to enable precision treatment for cancer, autoimmune diseases, and allergies.
Effectively designing such vaccines requires accurate prediction of proteasomal cleavage in order to ensure that the epitopes in the vaccine are presented to T cells by the major histocompatibility complex (MHC).
While direct identification of proteasomal cleavage \emph{in vitro} is cumbersome and low throughput, it is possible to implicitly infer cleavage events from the termini of MHC-presented epitopes, which can be detected in large amounts thanks to recent advances in high-throughput MHC ligandomics.
Inferring cleavage events in such a way provides an inherently noisy signal which can be tackled with new developments in the field of deep learning that supposedly make it possible to learn predictors from noisy labels.
Inspired by such innovations, we sought to modernize proteasomal cleavage predictors by benchmarking a wide range of recent methods, including LSTMs, transformers, CNNs, and denoising methods, on a recently introduced cleavage dataset.
We found that increasing model scale and complexity appeared to deliver limited performance gains, as several methods reached about 88.5\% AUC on C-terminal and 79.5\% AUC on N-terminal cleavage prediction.
This suggests that the noise and/or complexity of proteasomal cleavage and the subsequent biological processes of the antigen processing pathway are the major limiting factors for predictive performance rather than the specific modeling approach used.
While biological complexity can be tackled by more data and better models, noise and randomness inherently limit the maximum achievable predictive performance.
All our datasets and experiments are available at 
%\url{https://anonymous.4open.science/r/cleavage_prediction-E8FD}.
\url{https://github.com/ziegler-ingo/cleavage_prediction}.
\end{abstract}

\section{Introduction}
%The ability to include highly specific antigens makes epitope vaccines (EV) a promising avenue for precision medical treatment of conditions such as cancer, autoimmune diseases, and cancer~\citep{our-expert-opinion-paper}. 
Proteasomal cleavage digestion of antigens is a major step of the antigen processing pathway, as by cleaving proteins in smaller peptides it determines what may be subsequently presented by the major histocompatibility complex (MHC) to T cells, potentially triggering an immune response~\citep{antigen_pathway}.
Therefore, an important task for computational design of epitope vaccines (EV) is the prediction of this cleavage process, so that this information can be used by existing computational approaches~\citep{genev,jessev} to improve the efficacy of the vaccine.

Due to the difficulty of collecting large quantities of data \emph{in vitro}, proteasomal cleavage events are usually inferred implicitly from MHC ligandomics data~\citep{mass_spec} by matching eluted ligands to their progenitor protein to recover sequence information surrounding the terminals~\citep{netchop_first}.
This procedure, however, does not give an indication of which amino acid sequences \emph{cannot} result in a cleavage event, since missed cleavage sites are not observed in MHC ligands. Therefore, decoy negative samples are usually generated synthetically either by randomly shuffling the amino acids in a short window around the cleavage site or by considering artificial negative sites located around observed cleavage events~\citep{Calis2014}.  Even though such negative samples are not entirely reliable, the growing availability of this kind of data~\cite{iedb} spurred continuous development and improvement of proteasomal cleavage predictors~\cite{netchop_first,Kuttler2000,pcm,Nielsen_2005} which have been recently revised in light of new innovations in the deep learning field~\citep{netcleave,puuplclevage,Weeder2021,AmengualRigo2021}.

As a consequence of these developments, we implemented and tested several binary classification methods on a proteasomal cleavage prediction task, carefully benchmarking a wide choice of architectures, embeddings, and training regimes.

\section{Methods}
In this benchmark study we consider three main axis of variation: the initial embedding of amino acids, the neural architecture of the predictor, and their training regime via noise handling and data augmentations.

\subsection{Embedding}
%The concept of embeddings originally stems from natural language processing (NLP) and language models, but targeted adaptations for protein data have already been presented~\citep{ibtehaz2021application}.
%Embeddings constitute the first encoding step of a model.
The choice of embedding is crucial as it influences what intrinsic information a model can exploit for classification~\citep{ibtehaz2021application}; we thus consider various embeddings in our analysis, while keeping the base architecture equal.
Specifically, we analyze the performance of a randomly initialized embedding layer that is  optimized in conjunction with the loss function of the whole model, and the dedicated Prot2Vec~\citep{asgari2015continuous} embeddings trained with the well-established Word2Vec~\citep{mikolov_efficient_2013,mikolov_distributed_2013} algorithm.
Analogous to natural language, we design sequence embeddings by concatenating independently trained forward and backward amino acid representations of each input~\citep{DBLP:journals/corr/HeigoldNG16}.

%A current state-of-the-art approach in the NLP task of neural parts-of-speech tagging involves creating word representations by separately modeling character-level representations of an input token in both forward and backward directions.
%The concatenation of both encodings serves as the final embedding representation of a token and replaces the explicit token-level embedding layer~\citep{DBLP:journals/corr/HeigoldNG16}. 
%Analogous to natural language, we employ this technique to design sequence embeddings by concatenating independently trained forward and backward amino acid representations of each input.

Trainable tokenizers learn to form a given number of complex intra-token splits. This leads to a setting where the vocabulary size is now a tunable hyperparameter and 
thus has a direct impact on the size and quality of subsequently trained embedding representations. We extend our experiment with a vocabulary size \num{1000} and a 
vocabulary size \num{50000} version of the byte-level byte pair encoding~\citep[][BBPE]{sennrich_neural_2016}, as well as a vocabulary size \num{50000} version of the WordPiece~\citep[][WP]{schuster2012japanese} algorithm.

\subsection{Neural architectures}
\paragraph{Recurrent:}
Bidirectional long short-term memory networks (BiLSTM)~\citep{graves_framewise_2005} are well suited for a wide range of text classification tasks, thus
%As proteasomal cleavage prediction is performed on amino acid sequences in a sequence-to-one classification setting similar to text classification, 
we based nine of 12 model architectures around BiLSTMs.
The fundamental structure for our BiLSTMs is built around the architecture proposed by Ozols et al.~\citep{ijms22063071}, in which multiple sequential BiLSTMs are followed by a hidden and an output layer.
For eight of our nine BiLSTM-related experiments, we choose two sequential BiLSTMs, where sequence dimensionality is reduced by taking the maximum value of the depth-wise per-residue output of the last layer.
For the hidden layer, we used the Gaussian Error Linear Units (GELU)~\citep{hendrycks2016gaussian} activation function.
%The attention mechanism has also been applied to non-transformer architectures.
We additionally include an adjusted five BiLSTM version of a residual architecture between LSTM blocks, which aims to combat the shallow layer problem of deep LSTM architectures while also trying to improve the decoder quality with attention~\citep{liu2019attention}.

\paragraph{Transformers:}
Besides RNNs, the attention mechanism introduced by Vaswani et al. enabled a whole new architecture capable of processing sequences: the transformer~\citep{vaswani_attention_2017}.
%These complex multi-million and up to multi-billion parameter models feature an encoder and decoder system.
%Having access to the outputs of the encoder section of the model allows us to incorporate them as embeddings into our general model architecture.
We, therefore, integrated ProtTrans' T5-XL encoder-only model~\citep{9477085} featuring 1.2 billion parameters, as well as ESM2 transformer~\citep{Lin2022.07.20.500902} in its 150 million parameter version.
Additionally, we include a fine-tuning performance of ESM2 by adding a linear layer projection from its vocabulary-sized per-residue Roberta Language Model Head~\citep{liu_roberta_2019,doi:10.1073/pnas.2016239118} to our binary classification target.

\paragraph{Convolutional and Perceptron:}
We take the DeepCleave~\citep{10.1093/bioinformatics/btz721} attention-enhanced convolutional neural network~\citep[][CNN]{lecun_gradient-based_1998} architecture into our benchmark analysis.
Furthermore, stacking fully connected layers without any convolutional or recurrent features, e.g., in DeepCalpain~\citep{10.3389/fgene.2019.00715} or Terminitor~\citep{Yang710699}, has also been successfully applied to protein data.
As baseline, we include a single hidden layer perceptron~\citep{rumelhart1986learning} with Rectified Linear Units~\citep{DBLP:journals/corr/abs-1803-08375} as activation function into the analysis.

\subsection{Training}

\paragraph{Dataset:}
We used the dataset introduced in~\citep{puuplclevage}, which contains \num{229163} and \num{222181} N- and C-terminals cleavage sites respectively.
Each cleavage site is captured into a window comprising six amino acids to its left and four to its right, and is associated with six decoy negative samples obtained by considering the three residues preceding and following it, resulting in a total of \num{1434989} and \num{1419501} samples after deduplication for N- and C-terminals.
As the decoy negatives are situated in close proximity to real cleavage sites and due to the probabilistic nature of proteasomal cleavage, some of the negative samples are likely to be actual, unmeasured cleavage sites, and may influence the performance of predictors trained using such data.

\paragraph{Noisy labels:}
To reduce the impact of asymmetric label noise on the performance of our classifiers, we take five recent deep learning-specific denoising approaches into consideration: a noise adaptation layer, which attempts to learn the noise distribution in the data~\citep{goldberger2017training}, co-teaching, where two models are trained simultaneously by deciding for the respective other network which samples from a mini-batch to use for training~\citep{han2018co}, and co-teaching-plus~\citep{yu2019does}, which updates co-teaching with the disagreement learning approach of decoupling~\citep{malach2017decoupling}.
We additionally consider a joint training method with co-regularization (JoCoR)~\citep{wei2020combating} and DivideMix~\citep{Li2020DivideMix:} for benchmarking.
%JoCoR targets agreement learning by calculating a joint loss under co-regularization, which effectively reduces divergent predictions of two networks.
%It chooses the small-loss samples and updates the parameters of both networks simultaneously.
DivideMix is a holistic approach originally developed for computer vision and integrates multiple frameworks, such as co-teaching and MixMatch~\citep{berthelot2019mixmatch}, into one.
%Following the co-teaching approach, each network feeds the other network parts of the training dataset.
%In this case, however, the division of samples happens at the beginning of an epoch by fitting a Gaussian Mixture Model (GMM) ~\citep{Reynolds2009GaussianMM} and splitting the dataset at a chosen threshold.
%High-loss samples, as classified by the GMM, are treated as unlabeled and semi-supervised learning techniques based on MixMatch are followed.
As MixMatch builds upon MixUp~\citep{mixup}, which was developed for image data, we adjust it for sequential data by mixing up the embedded sequence representation~\citep{guo2019augmenting} instead of the pixel input in the data loading process.

\paragraph{Data augmentation:}
For all models, we apply data augmentation directly on the input sequences to combat overfitting and improve generalizability by masking a random amino acid per sequence as unknown~\citep{Shen2021.02.18.431877}.
All predictors except ESM2 fine-tuning use adaptive momentum~\citep{Kingma2015AdamAM} as their optimization technique, whereas ESM2 fine-tuning uses adaptive momentum with decoupled weight decay~\citep{loshchilov2017decoupled}.
All models without denoising techniques use (binary) cross-entropy loss~\citep{cox_regression_1958}, while all denoising models calculate dedicated losses.

\section{Experimental protocol}
\paragraph{Evaluation:}
As previously mentioned, some negative samples may actually result in a proteasomal cleavage event \emph{in vivo} due to the way these negative samples are generated.
For this reason, traditional binary classification metrics such as accuracy, precision, recall, etc. are misleading and model evaluation should instead be based on the AUC~\citep{Menon2015LearningFC}. 
We reserved a random 10\% of each terminal dataset as test dataset used for the final evaluation of the best hyperparameters. 

\paragraph{Hyperparameter optimization:}
%All model training takes place on one Nvidia RTX 3090. 
%To gather initial broad ranges of well-performing hyperparameters, we did manual try-out runs in a grid search approach.
%The final hyperparameter combinations were found using Ray Tune's~\citep{moritz2018ray} implementation of the asynchronous hyperband algorithm~\citep{li2020system} in a configuration with one bracket, a grace period of four, a reduction factor of two, and 30 sample runs.
%Models of group one evaluated the chosen hyperpameters in a 10-fold cross-validation, while models in group two performed 5-fold cross-validation.
%For models of group three, parameters were chosen manually according to initial try-out runs and the model with the best validation set performance was chosen for test set evaluation.
%For denoising methods, we used the same hyperparameters as the architecture without denoising and did not perform cross-validation.
%Denoising with DivideMix was only evaluated for the best-performing architecture as it increased training time by a factor of 8.5. 
Due to computational limitations, we split up the hyperparameter search into three priority groups: group one used Ray Tune's~\citep{moritz2018ray} implementation of the asynchronous hyperband algorithm~\citep{li2020system} and evaluated each configuration in a ten-folds cross-validation (CV), while for groups two and three we chose hyperparameters manually and evaluated each configuration with five-folds CV (group two) or a single run on a held-out validation set (group three).
We then used the best hyperparameter combination to train each architecture with all denoising methods, except for DivideMix where we only trained the overall best performing architecture due to computational limitations.
Information on the different architectures is in Appendix~\ref{apx:architectureinfos}, while the exact hyperband ranges and chosen hyperparameters for all models can be found in Appendix~\ref{apx:hyperparams}.

\section{Results}
\begin{figure}
  \centering
  \includegraphics[width=\textwidth]{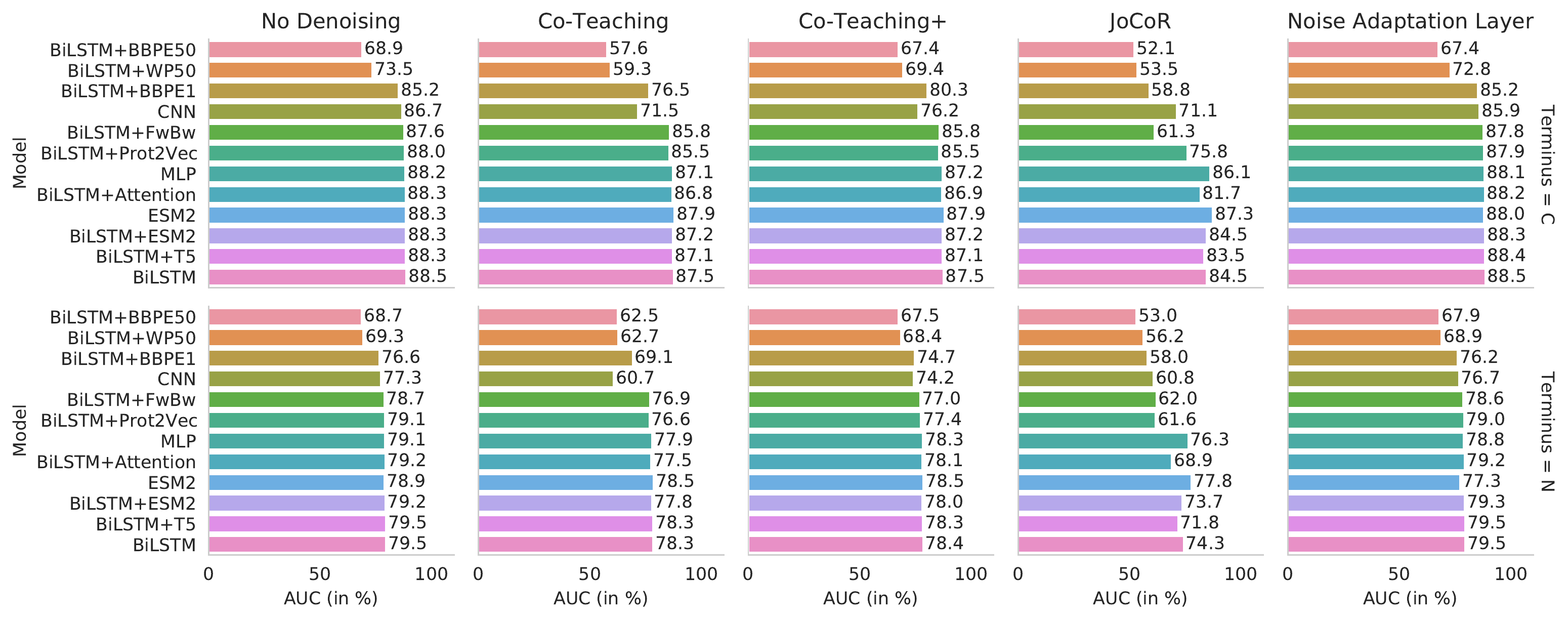}
  \caption{Model performances on C- and N-terminal}
  \label{fig:res}
\end{figure}

%Figure 1 shows the respective C- and N-terminal comparisons of the 12 studied model architectures across all discussed denoising approaches.
The overall best performing C-terminal model architecture as measured by AUC was the BiLSTM at 88.55\% without any denoising methods, while for the N-term, the BiLSTM+T5 with noise adaptation layer version narrowly outperformed the base BiLSTM version by 0.04 percentage points at a level of 79.54\% AUC (Figure~\ref{fig:res} and Appendix~\ref{apx:resultsnodenoise} and~\ref{apx:resultswithdenoise}).
If denoising techniques were applied, the noise adaptation layer consistently performed best for both the C- and N-terminal.
However, in 11 (10) of 12 models for the C-terminal (N-terminal), no denoising method resulted in superior results.
Co-teaching-plus dominated co-teaching along all (11) model architectures in the C-terminal (N-terminal).
JoCoR appeared to significantly hinder model performance in all architectures, whereas DivideMix also reduced the BiLSTM AUC score by around 2.4 percentage points in both terminals.
While the best-performing BiLSTM consisted of 4.6 million parameters, the MLP with \num{30,529} parameters only lacked 0.38 percentage points AUC behind and additionally beat the pre-trained Prot2Vec as well as DeepCleave architectures, both featuring around 16 million parameters in the C-terminal.
All transformer architectures in the ranges of 148 million (ESM2-based) and 1.2 billion (T5) parameters ranked behind the BiLSTM architecture but narrowly outperformed the MLP with AUC scores of around 88.32\%.

For the C-terminal, models including trainable tokenizer dropped to their worst-performing state compared to their fixed-vocabulary counterpart, especially when increasing the number of to-be-learned amino acid sub-string combinations. Whereas the BiLSTM with BBPE vocabulary size \num{1000} drops 3.3 percentage points to 85.25\% AUC, the same model architecture with \num{50000} learned sub-string combinations was only able to reach 68.92\% AUC. A similar but less severe pattern could be observed with WordPiece encodings, where the size \num{50000} vocabulary version reached 73.46\% AUC. If these models additionally featured denoising methods, the performance loss intensified up to a level of almost random-guessing (52.09\% AUC for \num{50000} BBPE and JoCoR). 

Interestingly, the N-terminal showed a different behavior for certain architecture combinations. BiLSTM+Attention and BiLSTM+Prot2Vec had significantly larger performance drop-offs from their best-performing model for JoCoR-denoising (10.7 and 18 percentage points, respectively) compared to the C-terminal (6.8 and 12.7 percentage points, respectively). On the other hand, the performance loss due to trainable tokenizers paired with JoCoR was less severe in the N-terminal (21.5, 26.5, 23.2 percentage points, respectively) for BBPE-\num{1000}, BBPE-\num{50000}, and WP-\num{50000} than in the C-terminal (29.7, 36.4, 35.1 percentage points, respectively).

Replacing the embedding layer with a forward-backward representation yielded comparable performance to the base BiLSTM architecture. Nonetheless, the base BiLSTM architecture was preferable as the additional forward-backward encoding steps increased training time by a factor of six.

\section{Conclusion}
Our benchmarking of various deep learning architectures for the task of proteasomal cleavage prediction has shown that several embedding techniques in combination with model architectures of vastly different scale and complexity can reach a performance of around 88.5\% AUC for C-terminals and 79.5\% AUC for N-terminals. Denoising techniques as well as trainable tokenizers appeared to offer limited to no, or even negative benefit. 
Such saturated results suggest that different modeling choices of architectures, embeddings, or training regimes are unlikely to yield significantly better predictive performance, and further efforts for proteasomal cleavage prediction should focus on a more comprehensive modeling of the antigen pathway.
Another possibility is that these biological processes are simply too noisy and random to allow more accurate predictions, in which case we may already be close to the boundary of what is possible to achieve.

\begin{ack}
%Use unnumbered first level headings for the acknowledgments. All acknowledgments
%go at the end of the paper before the list of references. Moreover, you are required to declare
%funding (financial activities supporting the submitted work) and competing interests (related financial activities outside the submitted work).
%More information about this disclosure can be found at: \url{https://neurips.cc/Conferences/2022/PaperInformation/FundingDisclosure}.
%Do {\bf not} include this section in the anonymized submission, only in the final paper. You can use the \texttt{ack} environment provided in the style file to autmoatically hide this section in the anonymized submission.

E. D. was supported by the Helmholtz Association under the joint research school ``Munich School for Data Science - MUDS`` (Award Number HIDSS-0006).
B. S. acknowledges financial support by the Postdoctoral Fellowship Program of the Helmholtz Zentrum M{\"u}nchen.
\end{ack}

{
\small
\bibliographystyle{unsrtnat}
\bibliography{main}

\begin{thebibliography}{52}
\providecommand{\natexlab}[1]{#1}
\providecommand{\url}[1]{\texttt{#1}}
\expandafter\ifx\csname urlstyle\endcsname\relax
  \providecommand{\doi}[1]{doi: #1}\else
  \providecommand{\doi}{doi: \begingroup \urlstyle{rm}\Url}\fi

\bibitem[Blum et~al.(2013)Blum, Wearsch, and Cresswell]{antigen_pathway}
Janice~S. Blum, Pamela~A. Wearsch, and Peter Cresswell.
\newblock Pathways of antigen processing.
\newblock \emph{Annual Review of Immunology}, 31\penalty0 (1):\penalty0
  443--473, March 2013.
\newblock \doi{10.1146/annurev-immunol-032712-095910}.

\bibitem[Dorigatti and Schubert(2020{\natexlab{a}})]{genev}
Emilio Dorigatti and Benjamin Schubert.
\newblock Graph-theoretical formulation of the generalized epitope-based
  vaccine design problem.
\newblock \emph{{PLOS} Computational Biology}, 16\penalty0 (10):\penalty0
  e1008237, October 2020{\natexlab{a}}.
\newblock \doi{10.1371/journal.pcbi.1008237}.

\bibitem[Dorigatti and Schubert(2020{\natexlab{b}})]{jessev}
Emilio Dorigatti and Benjamin Schubert.
\newblock Joint epitope selection and spacer design for string-of-beads
  vaccines.
\newblock \emph{Bioinformatics}, 36\penalty0 (Supplement{\_}2):\penalty0
  i643--i650, December 2020{\natexlab{b}}.
\newblock \doi{10.1093/bioinformatics/btaa790}.

\bibitem[Purcell et~al.(2019)Purcell, Ramarathinam, and Ternette]{mass_spec}
Anthony~W. Purcell, Sri~H. Ramarathinam, and Nicola Ternette.
\newblock Mass spectrometry{\textendash}based identification of {MHC}-bound
  peptides for immunopeptidomics.
\newblock \emph{Nat Protoc}, 14\penalty0 (6):\penalty0 1687--1707, may 2019.
\newblock \doi{10.1038/s41596-019-0133-y}.

\bibitem[Ke{\c{s}}mir et~al.(2002)Ke{\c{s}}mir, Nussbaum, Schild, Detours, and
  Brunak]{netchop_first}
Can Ke{\c{s}}mir, Alexander~K. Nussbaum, Hansj\"{o}rg Schild, Vincent Detours,
  and S{\o}ren Brunak.
\newblock Prediction of proteasome cleavage motifs by neural networks.
\newblock \emph{Protein Engineering, Design and Selection}, 15\penalty0
  (4):\penalty0 287--296, April 2002.
\newblock \doi{10.1093/protein/15.4.287}.

\bibitem[Calis et~al.(2014)Calis, Reinink, Keller, Kloetzel, and
  Ke{\c{s}}mir]{Calis2014}
Jorg J.~A. Calis, Peter Reinink, Christin Keller, Peter~M. Kloetzel, and Can
  Ke{\c{s}}mir.
\newblock Role of peptide processing predictions in t cell epitope
  identification: contribution of different prediction programs.
\newblock \emph{Immunogenetics}, 67\penalty0 (2):\penalty0 85--93, December
  2014.
\newblock \doi{10.1007/s00251-014-0815-0}.
\newblock URL \url{https://doi.org/10.1007/s00251-014-0815-0}.

\bibitem[Vita et~al.(2018)Vita, Mahajan, Overton, Dhanda, Martini, Cantrell,
  Wheeler, Sette, and Peters]{iedb}
Randi Vita, Swapnil Mahajan, James~A Overton, Sandeep~Kumar Dhanda, Sheridan
  Martini, Jason~R Cantrell, Daniel~K Wheeler, Alessandro Sette, and Bjoern
  Peters.
\newblock {The Immune Epitope Database (IEDB): 2018 update}.
\newblock \emph{Nucleic Acids Research}, 47\penalty0 (D1):\penalty0 D339--D343,
  10 2018.
\newblock ISSN 0305-1048.
\newblock \doi{10.1093/nar/gky1006}.

\bibitem[Kuttler et~al.(2000)Kuttler, Nussbaum, Dick, Rammensee, Schild, and
  Hadeler]{Kuttler2000}
Christina Kuttler, Alexander~K Nussbaum, Tobias~P Dick, Hans-Georg Rammensee,
  Hansj\"{o}rg Schild, and Karl-Peter Hadeler.
\newblock An algorithm for the prediction of proteasomal cleavages.
\newblock \emph{Journal of Molecular Biology}, 298\penalty0 (3):\penalty0
  417--429, May 2000.
\newblock \doi{10.1006/jmbi.2000.3683}.

\bibitem[D\"{o}nnes and Kohlbacher(2005)]{pcm}
Pierre D\"{o}nnes and Oliver Kohlbacher.
\newblock Integrated modeling of the major events in the {MHC} class i antigen
  processing pathway.
\newblock \emph{Protein Science}, 14\penalty0 (8):\penalty0 2132--2140, August
  2005.
\newblock \doi{10.1110/ps.051352405}.

\bibitem[Nielsen et~al.(2005)Nielsen, Lundegaard, Lund, and
  Ke{\c{s}}mir]{Nielsen_2005}
Morten Nielsen, Claus Lundegaard, Ole Lund, and Can Ke{\c{s}}mir.
\newblock The role of the proteasome in generating cytotoxic t-cell epitopes:
  insights obtained from improved predictions of proteasomal cleavage.
\newblock \emph{Immunogenetics}, 57\penalty0 (1-2):\penalty0 33--41, mar 2005.
\newblock \doi{10.1007/s00251-005-0781-7}.

\bibitem[Amengual-Rigo and Guallar(2021{\natexlab{a}})]{netcleave}
Pep Amengual-Rigo and Victor Guallar.
\newblock {NetCleave}: an open-source algorithm for predicting c-terminal
  antigen processing for {MHC}-i and {MHC}-{II}.
\newblock \emph{Scientific Reports}, 11\penalty0 (1), June 2021{\natexlab{a}}.
\newblock \doi{10.1038/s41598-021-92632-y}.

\bibitem[Dorigatti et~al.(2022)Dorigatti, Bischl, and Schubert]{puuplclevage}
Emilio Dorigatti, Bernd Bischl, and Benjamin Schubert.
\newblock Improved proteasomal cleavage prediction with positive-unlabeled
  learning.
\newblock \emph{arXiv preprint arXiv:2209.07527}, 2022.

\bibitem[Weeder et~al.(2021)Weeder, Wood, Li, Nellore, and
  Thompson]{Weeder2021}
Benjamin~R Weeder, Mary~A Wood, Ellysia Li, Abhinav Nellore, and Reid~F
  Thompson.
\newblock pepsickle rapidly and accurately predicts proteasomal cleavage sites
  for improved neoantigen identification.
\newblock \emph{Bioinformatics}, 37\penalty0 (21):\penalty0 3723--3733,
  September 2021.
\newblock \doi{10.1093/bioinformatics/btab628}.
\newblock URL \url{https://doi.org/10.1093/bioinformatics/btab628}.

\bibitem[Amengual-Rigo and Guallar(2021{\natexlab{b}})]{AmengualRigo2021}
Pep Amengual-Rigo and Victor Guallar.
\newblock {NetCleave}: an open-source algorithm for predicting c-terminal
  antigen processing for {MHC}-i and {MHC}-{II}.
\newblock \emph{Scientific Reports}, 11\penalty0 (1), June 2021{\natexlab{b}}.
\newblock \doi{10.1038/s41598-021-92632-y}.
\newblock URL \url{https://doi.org/10.1038/s41598-021-92632-y}.

\bibitem[Ibtehaz and Kihara(2021)]{ibtehaz2021application}
Nabil Ibtehaz and Daisuke Kihara.
\newblock Application of sequence embedding in protein sequence-based
  predictions.
\newblock \emph{arXiv preprint arXiv:2110.07609}, 2021.

\bibitem[Asgari and Mofrad(2015)]{asgari2015continuous}
Ehsaneddin Asgari and Mohammad~RK Mofrad.
\newblock Continuous distributed representation of biological sequences for
  deep proteomics and genomics.
\newblock \emph{PloS one}, 10\penalty0 (11):\penalty0 e0141287, 2015.

\bibitem[Mikolov et~al.(2013{\natexlab{a}})Mikolov, Chen, Corrado, and
  Dean]{mikolov_efficient_2013}
Tomas Mikolov, Kai Chen, Greg Corrado, and Jeffrey Dean.
\newblock Efficient {Estimation} of {Word} {Representations} in {Vector}
  {Space}.
\newblock \emph{arXiv:1301.3781 [cs]}, September 2013{\natexlab{a}}.
\newblock URL \url{http://arxiv.org/abs/1301.3781}.
\newblock arXiv: 1301.3781.

\bibitem[Mikolov et~al.(2013{\natexlab{b}})Mikolov, Sutskever, Chen, Corrado,
  and Dean]{mikolov_distributed_2013}
Tomas Mikolov, Ilya Sutskever, Kai Chen, Greg Corrado, and Jeffrey Dean.
\newblock Distributed {Representations} of {Words} and {Phrases} and their
  {Compositionality}.
\newblock \emph{arXiv:1310.4546 [cs, stat]}, October 2013{\natexlab{b}}.
\newblock URL \url{http://arxiv.org/abs/1310.4546}.
\newblock arXiv: 1310.4546.

\bibitem[Heigold et~al.(2016)Heigold, Neumann, and van
  Genabith]{DBLP:journals/corr/HeigoldNG16}
Georg Heigold, Guenter Neumann, and Josef van Genabith.
\newblock Neural morphological tagging from characters for morphologically rich
  languages.
\newblock \emph{CoRR}, abs/1606.06640, 2016.
\newblock URL \url{http://arxiv.org/abs/1606.06640}.

\bibitem[Sennrich et~al.(2016)Sennrich, Haddow, and
  Birch]{sennrich_neural_2016}
Rico Sennrich, Barry Haddow, and Alexandra Birch.
\newblock Neural {Machine} {Translation} of {Rare} {Words} with {Subword}
  {Units}.
\newblock In \emph{Proceedings of the 54th {Annual} {Meeting} of the
  {Association} for {Computational} {Linguistics} ({Volume} 1: {Long}
  {Papers})}, pages 1715--1725, Berlin, Germany, 2016. Association for
  Computational Linguistics.
\newblock \doi{10.18653/v1/P16-1162}.
\newblock URL \url{http://aclweb.org/anthology/P16-1162}.

\bibitem[Schuster and Nakajima(2012)]{schuster2012japanese}
Mike Schuster and Kaisuke Nakajima.
\newblock Japanese and korean voice search.
\newblock In \emph{2012 IEEE international conference on acoustics, speech and
  signal processing (ICASSP)}, pages 5149--5152. IEEE, 2012.

\bibitem[Graves and Schmidhuber(2005)]{graves_framewise_2005}
Alex Graves and Jürgen Schmidhuber.
\newblock Framewise phoneme classification with bidirectional {LSTM} and other
  neural network architectures.
\newblock \emph{Neural Networks}, 18\penalty0 (5-6):\penalty0 602--610, July
  2005.
\newblock ISSN 08936080.
\newblock \doi{10.1016/j.neunet.2005.06.042}.
\newblock URL
  \url{https://linkinghub.elsevier.com/retrieve/pii/S0893608005001206}.

\bibitem[Ozols et~al.(2021)Ozols, Eckersley, Platt, Stewart-McGuinness,
  Hibbert, Revote, Li, Griffiths, Watson, Song, Bell, and
  Sherratt]{ijms22063071}
Matiss Ozols, Alexander Eckersley, Christopher~I. Platt, Callum
  Stewart-McGuinness, Sarah~A. Hibbert, Jerico Revote, Fuyi Li, Christopher
  E.~M. Griffiths, Rachel E.~B. Watson, Jiangning Song, Mike Bell, and
  Michael~J. Sherratt.
\newblock Predicting proteolysis in complex proteomes using deep learning.
\newblock \emph{International Journal of Molecular Sciences}, 22\penalty0 (6),
  2021.
\newblock ISSN 1422-0067.
\newblock \doi{10.3390/ijms22063071}.
\newblock URL \url{https://www.mdpi.com/1422-0067/22/6/3071}.

\bibitem[Hendrycks and Gimpel(2016)]{hendrycks2016gaussian}
Dan Hendrycks and Kevin Gimpel.
\newblock Gaussian error linear units (gelus).
\newblock \emph{arXiv preprint arXiv:1606.08415}, 2016.

\bibitem[Liu and Gong(2019)]{liu2019attention}
Jiale Liu and Xinqi Gong.
\newblock Attention mechanism enhanced lstm with residual architecture and its
  application for protein-protein interaction residue pairs prediction.
\newblock \emph{BMC bioinformatics}, 20\penalty0 (1):\penalty0 1--11, 2019.

\bibitem[Vaswani et~al.(2017)Vaswani, Shazeer, Parmar, Uszkoreit, Jones, Gomez,
  Kaiser, and Polosukhin]{vaswani_attention_2017}
Ashish Vaswani, Noam Shazeer, Niki Parmar, Jakob Uszkoreit, Llion Jones,
  Aidan~N. Gomez, Lukasz Kaiser, and Illia Polosukhin.
\newblock Attention {Is} {All} {You} {Need}.
\newblock \emph{arXiv:1706.03762 [cs]}, December 2017.
\newblock URL \url{http://arxiv.org/abs/1706.03762}.
\newblock arXiv: 1706.03762.

\bibitem[Elnaggar et~al.(2022)Elnaggar, Heinzinger, Dallago, Rehawi, Wang,
  Jones, Gibbs, Feher, Angerer, Steinegger, Bhowmik, and Rost]{9477085}
Ahmed Elnaggar, Michael Heinzinger, Christian Dallago, Ghalia Rehawi, Yu~Wang,
  Llion Jones, Tom Gibbs, Tamas Feher, Christoph Angerer, Martin Steinegger,
  Debsindhu Bhowmik, and Burkhard Rost.
\newblock Prottrans: Toward understanding the language of life through
  self-supervised learning.
\newblock \emph{IEEE Transactions on Pattern Analysis and Machine
  Intelligence}, 44\penalty0 (10):\penalty0 7112--7127, 2022.
\newblock \doi{10.1109/TPAMI.2021.3095381}.

\bibitem[Lin et~al.(2022)Lin, Akin, Rao, Hie, Zhu, Lu, Santos~Costa,
  Fazel-Zarandi, Sercu, Candido, and Rives]{Lin2022.07.20.500902}
Zeming Lin, Halil Akin, Roshan Rao, Brian Hie, Zhongkai Zhu, Wenting Lu,
  Allan~dos Santos~Costa, Maryam Fazel-Zarandi, Tom Sercu, Sal Candido, and
  Alexander Rives.
\newblock Language models of protein sequences at the scale of evolution enable
  accurate structure prediction.
\newblock \emph{bioRxiv}, 2022.
\newblock \doi{10.1101/2022.07.20.500902}.
\newblock URL
  \url{https://www.biorxiv.org/content/early/2022/07/21/2022.07.20.500902}.

\bibitem[Liu et~al.(2019{\natexlab{a}})Liu, Ott, Goyal, Du, Joshi, Chen, Levy,
  Lewis, Zettlemoyer, and Stoyanov]{liu_roberta_2019}
Yinhan Liu, Myle Ott, Naman Goyal, Jingfei Du, Mandar Joshi, Danqi Chen, Omer
  Levy, Mike Lewis, Luke Zettlemoyer, and Veselin Stoyanov.
\newblock {RoBERTa}: {A} {Robustly} {Optimized} {BERT} {Pretraining}
  {Approach}.
\newblock \emph{arXiv:1907.11692 [cs]}, July 2019{\natexlab{a}}.
\newblock URL \url{http://arxiv.org/abs/1907.11692}.
\newblock arXiv: 1907.11692.

\bibitem[Rives et~al.(2021)Rives, Meier, Sercu, Goyal, Lin, Liu, Guo, Ott,
  Zitnick, Ma, and Fergus]{doi:10.1073/pnas.2016239118}
Alexander Rives, Joshua Meier, Tom Sercu, Siddharth Goyal, Zeming Lin, Jason
  Liu, Demi Guo, Myle Ott, C.~Lawrence Zitnick, Jerry Ma, and Rob Fergus.
\newblock Biological structure and function emerge from scaling unsupervised
  learning to 250 million protein sequences.
\newblock \emph{Proceedings of the National Academy of Sciences}, 118\penalty0
  (15):\penalty0 e2016239118, 2021.
\newblock \doi{10.1073/pnas.2016239118}.
\newblock URL \url{https://www.pnas.org/doi/abs/10.1073/pnas.2016239118}.

\bibitem[Li et~al.(2019)Li, Chen, Leier, Marquez-Lago, Liu, Wang, Revote,
  Smith, Akutsu, Webb, Kurgan, and Song]{10.1093/bioinformatics/btz721}
Fuyi Li, Jinxiang Chen, André Leier, Tatiana Marquez-Lago, Quanzhong Liu,
  Yanze Wang, Jerico Revote, A~Ian Smith, Tatsuya Akutsu, Geoffrey~I Webb,
  Lukasz Kurgan, and Jiangning Song.
\newblock {DeepCleave: a deep learning predictor for caspase and matrix
  metalloprotease substrates and cleavage sites}.
\newblock \emph{Bioinformatics}, 36\penalty0 (4):\penalty0 1057--1065, 09 2019.
\newblock ISSN 1367-4803.
\newblock \doi{10.1093/bioinformatics/btz721}.
\newblock URL \url{https://doi.org/10.1093/bioinformatics/btz721}.

\bibitem[LeCun et~al.(1998)LeCun, Bottou, Bengio, and
  Haffner]{lecun_gradient-based_1998}
Yann LeCun, Leon Bottou, Yoshua Bengio, and Patrick Haffner.
\newblock Gradient-{Based} {Learning} {Applied} to {Document} {Recognition}.
\newblock \emph{Proc. of the IEEE}, pages 1--46, 1998.

\bibitem[Liu et~al.(2019{\natexlab{b}})Liu, Yu, Dong, Zhao, Liu, Zhang, Li, Du,
  and Cheng]{10.3389/fgene.2019.00715}
Ze-Xian Liu, Kai Yu, Jingsi Dong, Linhong Zhao, Zekun Liu, Qingfeng Zhang,
  Shihua Li, Yimeng Du, and Han Cheng.
\newblock Precise prediction of calpain cleavage sites and their aberrance
  caused by mutations in cancer.
\newblock \emph{Frontiers in Genetics}, 10, 2019{\natexlab{b}}.
\newblock ISSN 1664-8021.
\newblock \doi{10.3389/fgene.2019.00715}.
\newblock URL
  \url{https://www.frontiersin.org/articles/10.3389/fgene.2019.00715}.

\bibitem[Yang et~al.(2020)Yang, Li, Nip, Warren, and Birol]{Yang710699}
Chen Yang, Chenkai Li, Ka~Ming Nip, Ren{\'e}~L Warren, and Inanc Birol.
\newblock Terminitor: Cleavage site prediction using deep learning models.
\newblock \emph{bioRxiv}, 2020.
\newblock \doi{10.1101/710699}.
\newblock URL \url{https://www.biorxiv.org/content/early/2020/04/23/710699}.

\bibitem[Rumelhart et~al.(1986)Rumelhart, Hinton, and
  Williams]{rumelhart1986learning}
David~E Rumelhart, Geoffrey~E Hinton, and Ronald~J Williams.
\newblock Learning representations by back-propagating errors.
\newblock \emph{nature}, 323\penalty0 (6088):\penalty0 533--536, 1986.

\bibitem[Agarap(2018)]{DBLP:journals/corr/abs-1803-08375}
Abien~Fred Agarap.
\newblock Deep learning using rectified linear units (relu).
\newblock \emph{CoRR}, abs/1803.08375, 2018.
\newblock URL \url{http://arxiv.org/abs/1803.08375}.

\bibitem[Goldberger and Ben-Reuven(2017)]{goldberger2017training}
Jacob Goldberger and Ehud Ben-Reuven.
\newblock Training deep neural-networks using a noise adaptation layer.
\newblock In \emph{International Conference on Learning Representations}, 2017.
\newblock URL \url{https://openreview.net/forum?id=H12GRgcxg}.

\bibitem[Han et~al.(2018)Han, Yao, Yu, Niu, Xu, Hu, Tsang, and
  Sugiyama]{han2018co}
Bo~Han, Quanming Yao, Xingrui Yu, Gang Niu, Miao Xu, Weihua Hu, Ivor Tsang, and
  Masashi Sugiyama.
\newblock Co-teaching: Robust training of deep neural networks with extremely
  noisy labels.
\newblock \emph{Advances in neural information processing systems}, 31, 2018.

\bibitem[Yu et~al.(2019)Yu, Han, Yao, Niu, Tsang, and Sugiyama]{yu2019does}
Xingrui Yu, Bo~Han, Jiangchao Yao, Gang Niu, Ivor Tsang, and Masashi Sugiyama.
\newblock How does disagreement help generalization against label corruption?
\newblock In \emph{International Conference on Machine Learning}, pages
  7164--7173. PMLR, 2019.

\bibitem[Malach and Shalev-Shwartz(2017)]{malach2017decoupling}
Eran Malach and Shai Shalev-Shwartz.
\newblock Decoupling" when to update" from" how to update".
\newblock \emph{Advances in neural information processing systems}, 30, 2017.

\bibitem[Wei et~al.(2020)Wei, Feng, Chen, and An]{wei2020combating}
Hongxin Wei, Lei Feng, Xiangyu Chen, and Bo~An.
\newblock Combating noisy labels by agreement: A joint training method with
  co-regularization.
\newblock In \emph{Proceedings of the IEEE/CVF Conference on Computer Vision
  and Pattern Recognition}, pages 13726--13735, 2020.

\bibitem[Li et~al.(2020{\natexlab{a}})Li, Socher, and Hoi]{Li2020DivideMix:}
Junnan Li, Richard Socher, and Steven~C.H. Hoi.
\newblock Dividemix: Learning with noisy labels as semi-supervised learning.
\newblock In \emph{International Conference on Learning Representations},
  2020{\natexlab{a}}.
\newblock URL \url{https://openreview.net/forum?id=HJgExaVtwr}.

\bibitem[Berthelot et~al.(2019)Berthelot, Carlini, Goodfellow, Papernot,
  Oliver, and Raffel]{berthelot2019mixmatch}
David Berthelot, Nicholas Carlini, Ian Goodfellow, Nicolas Papernot, Avital
  Oliver, and Colin~A Raffel.
\newblock Mixmatch: A holistic approach to semi-supervised learning.
\newblock \emph{Advances in neural information processing systems}, 32, 2019.

\bibitem[Zhang et~al.(2018)Zhang, Cisse, Dauphin, and Lopez-Paz]{mixup}
Hongyi Zhang, Moustapha Cisse, Yann~N. Dauphin, and David Lopez-Paz.
\newblock Mixup: Beyond empirical risk minimization.
\newblock \emph{International Conference on Learning Representations}, 2018.

\bibitem[Guo et~al.(2019)Guo, Mao, and Zhang]{guo2019augmenting}
Hongyu Guo, Yongyi Mao, and Richong Zhang.
\newblock Augmenting data with mixup for sentence classification: An empirical
  study.
\newblock \emph{arXiv preprint arXiv:1905.08941}, 2019.

\bibitem[Shen et~al.(2021)Shen, Price, Bahadori, and
  Seeger]{Shen2021.02.18.431877}
Hongyu Shen, Layne~C. Price, Taha Bahadori, and Franziska Seeger.
\newblock Improving generalizability of protein sequence models with data
  augmentations.
\newblock \emph{bioRxiv}, 2021.
\newblock \doi{10.1101/2021.02.18.431877}.
\newblock URL
  \url{https://www.biorxiv.org/content/early/2021/02/18/2021.02.18.431877}.

\bibitem[Kingma and Ba(2015)]{Kingma2015AdamAM}
Diederik~P. Kingma and Jimmy Ba.
\newblock Adam: A method for stochastic optimization.
\newblock \emph{CoRR}, abs/1412.6980, 2015.

\bibitem[Loshchilov and Hutter(2017)]{loshchilov2017decoupled}
Ilya Loshchilov and Frank Hutter.
\newblock Decoupled weight decay regularization.
\newblock \emph{arXiv preprint arXiv:1711.05101}, 2017.

\bibitem[Cox(1958)]{cox_regression_1958}
D.R. Cox.
\newblock The {Regression} {Analysis} of binary sequences.
\newblock \emph{Journal of the Royal Statistical Society: Series B
  (Methodological)}, 2\penalty0 (2), 1958.
\newblock ISSN 00359246.

\bibitem[Menon et~al.(2015)Menon, Rooyen, Ong, and
  Williamson]{Menon2015LearningFC}
Aditya Menon, Brendan~Van Rooyen, Cheng~Soon Ong, and Bob Williamson.
\newblock Learning from corrupted binary labels via class-probability
  estimation.
\newblock In Francis Bach and David Blei, editors, \emph{Proceedings of the
  32nd International Conference on Machine Learning}, volume~37 of
  \emph{Proceedings of Machine Learning Research}, pages 125--134, Lille,
  France, 07--09 Jul 2015. PMLR.

\bibitem[Moritz et~al.(2018)Moritz, Nishihara, Wang, Tumanov, Liaw, Liang,
  Elibol, Yang, Paul, Jordan, et~al.]{moritz2018ray}
Philipp Moritz, Robert Nishihara, Stephanie Wang, Alexey Tumanov, Richard Liaw,
  Eric Liang, Melih Elibol, Zongheng Yang, William Paul, Michael~I Jordan,
  et~al.
\newblock Ray: A distributed framework for emerging $\{$AI$\}$ applications.
\newblock In \emph{13th USENIX Symposium on Operating Systems Design and
  Implementation (OSDI 18)}, pages 561--577, 2018.

\bibitem[Li et~al.(2020{\natexlab{b}})Li, Jamieson, Rostamizadeh, Gonina,
  Ben-Tzur, Hardt, Recht, and Talwalkar]{li2020system}
Liam Li, Kevin Jamieson, Afshin Rostamizadeh, Ekaterina Gonina, Jonathan
  Ben-Tzur, Moritz Hardt, Benjamin Recht, and Ameet Talwalkar.
\newblock A system for massively parallel hyperparameter tuning.
\newblock \emph{Proceedings of Machine Learning and Systems}, 2:\penalty0
  230--246, 2020{\natexlab{b}}.

\end{thebibliography}
}

%%%%%%%%%%%%%%%%%%%%%%%%%%%%%%%%%%%%%%%%%%%%%%%%%%%%%%%%%%%%

\newpage
\appendix

\section{Appendix}

% comment following two tables
\begin{comment}
\begin{table}[H]
  \caption{Models and achieved AUC on C-terminal}
  \label{models-auc-c}
  \centering
  \scriptsize
  \begin{tabular}{llll}
    \toprule
    \textbf{Models}  & \textbf{AUC} &  \textbf{Models}  & \textbf{AUC} \\
    \midrule
    BiLSTM  & 88.55 $\pm$ 0.12 & ESM2FineTuned & 88.32 $\pm$  0.16\\
    BiLSTMAttention & 88.28 $\pm$  0.08 & T5BiLSTM & 88.32 $\pm$  0.05\\
    BiLSTMProt2Vec & 87.99 $\pm$  0.14 & BiLSTMbbpe1k & 85.25 \\
    CNNAttention & 86.66 $\pm$  0.17 & BiLSTMbbpe50k & 68.92 \\
    MLP & 88.17 $\pm$  0.11 & BiLSTMbwp & 73.46\\
    ESM2EmbedBiLSTM & 88.34 $\pm$  0.05 & FwBwBiLSTM & 87.59\\
    \bottomrule
  \end{tabular}
\end{table}

\begin{table}[H]
  \caption{The AUC of BiLSTM with denoising on C-terminal}
  \label{bilstm-denoising}
  \centering
  \scriptsize
  \begin{tabular}{ll}
    \toprule
    \textbf{BiLSTM with denoising methods}  & \textbf{AUC}  \\
    \midrule
    Co-Teaching  & 87.50 \\
    Co-Teaching+ & 87.50\\
    JoCoR & 84.53 \\
    DivideMix & 86.25 \\
    Noise Adaptation Layer & 88.49\\
    \bottomrule
  \end{tabular}
\end{table}
\end{comment}

\subsection{Architecture information}
\label{apx:architectureinfos}
\begin{table}[H]
  \caption{Number of parameters and training time for each model, without considering denoising (see later tables for this)}
  \label{arch-info}
  \centering
  \scriptsize
  \begin{tabular}{lccrr}
    \toprule
    \textbf{Models} & \textbf{Time (s/epoch)} & \textbf{Epochs} & \textbf{Parameters} & \textbf{Trainable} \\
    \cmidrule(r){1-1}\cmidrule(r){2-3}\cmidrule(r){4-5}
    BiLSTM  &  25 & 15 &    \num{4655984} &\num{4655984}\\
    BiLSTM+Attention   & ~20 & 20 & \num{1632391} &\num{1632391} \\
    BiLSTM+Prot2Vec   &  ~20 &    15&    \num{16009371}    &\num{5830371} \\
    CNN   &  ~45 &    60    &\num{16084198}    &\num{16084198} \\
    MLP   &  ~4 &    30    & \num{30529} &\num{30529} \\
    BiLSTM+ESM2  &  ~330     & 10 &    \num{152998267} &\num{4858113} \\
    ESM2   &  ~900 &    3    & \num{148140188}     & \num{148140188} \\
    BiLSTM+T5  &  ~780 &    10 & \num{1214572801}  & \num{6430977}\\
    BiLSTM+BBPE1   &  ~13 & 15 & \num{5319409}&   \num{5319409} \\
    BiLSTM+BBPE50   &  ~12 & 15 & \num{12669409}&   \num{12669409}\\
    BiLSTM+WP50  &  ~16 & 15 & \num{12669409} &  \num{12669409}\\
    BiLSTM+FwBw   &  ~120 & 15 & \num{4315369}&  \num{4315369} \\
    \bottomrule
  \end{tabular}
  %{\raggedright \hspace{2.5mm} Note: The subscript number at the end of each model name indicates its priority group. \par}
\end{table}

%%%%HYPERPARAMETERS%%%%%%%%%%%%%%%%%%%%%%%%%%%%%%%%%%%%%%%%%%%%%%%%%%%%%%%%
%%%%%%%%%%%%%%%%%%%%%%%%%%%%%%%%%%%%%%%%%%%%%%%%%%%%%%%%
\subsection{Hyperparameters}
\label{apx:hyperparams}

\begin{table}[H]
  \caption{BiLSTM}
  \label{bilstm-hyperparams}
  \centering
  \scriptsize
  \begin{tabular}{lccc}
    \toprule
\multirow{2}*{\textbf{Hyperparameter}}  & \textbf{Range}  & \multicolumn{2}{c}{\textbf{Final value}} \\
& (Uniformly random choice) & \textbf{C-terminal} & \textbf{N-terminal}\\
    \midrule
%LR Reduction Factor &    2 &    - &    -\\
%Grace Period &    4 &    - &    -\\
%Sample Runs     & 30&     -&    -\\
%Parameters    &    4,655,984 &    5,096,135\\
%Trainable Parameters &        4,655,984    & 5,096,135\\
% stuff above isn't hyperparams, maybe there's a better place of putting this (eg another table with param no.)
%\cmidrule
Epochs trained &    $\leq$ 25 &    15 &    15\\
%Time (s/epoch) w/o denoising &    - &    25 &    25 \\
%Learning Rate &    tune.choice([5e-5, 1e-4, 3e-4])    & 3e-04 &    3e-04\\
Learning rate &    $\{5\times10^{-5}, 10^{-4}, 3\times10^{-4}\}$    & $3\times10^{-4}$  &    $3\times10^{-4}$ \\
%Dropout Rate &    tune.quniform(0.45, 0.52, 0.01)    & 0.5 &    0.5\\
Dropout rate &    $\{0.45,0.46,\ldots,0.51,0.52\}$    & 0.5 &    0.5\\
%Linear Layer Size 1    & tune.randint(120, 181) &    164    179\\
Linear layer size    & $[120,181)$ &    164    & 179\\
%Linear Layer Size 2    & - &    1 &    1\\
%Embedding Dimensions &    tune.randint(50, 201) &    91 &    76\\
Embedding dimension &    $[50,201)$ &    91 &    76\\
%LSTM Size 1     & tune.randint(220, 281) &    228    & 252\\
LSTM size 1     & $[220,281)$ &    228    & 252\\
%LSTM Size 2    & tune.randint(450, 520) &    506    & 518\\
LSTM size 2    & $[450,520)$ &    506    & 518\\
    \bottomrule
  \end{tabular}
\end{table}

\begin{table}[H]
  \caption{BiLSTM+Attention}
  \label{BiLSTMAttention-hyperparams}
  \centering
  \scriptsize
  \begin{tabular}{lccc}
    \toprule
    \multirow{2}*{\textbf{Hyperparameter}}  & \textbf{Range}  & \multicolumn{2}{c}{\textbf{Final value}} \\
    & (Uniformly random choice) & \textbf{C-terminal} & \textbf{N-terminal}\\
    \midrule
    Epochs trained &    $\leq$ 25 &    20 &    20\\
    Learning rate &    $\{3\times10^{-5}, 5\times10^{-5}, 8\times10^{-5},10^{-4}\}$    & $10^{-4}$ &    $10^{-4}$\\
    Dropout rate &    $\{0.45,0.46,\ldots,0.51,0.52\}$    & 0.5 &    0.5\\
    Linear layer size    & $[100,181)$ &    147    & 150\\
    Embedding dimension &    $\{120,124,\ldots,216,220\}$ &    216 &    216\\
    LSTM size     & $[64,131)$ &    108    & 111\\
    Attention heads    & $\{1,2,4\}$ &    4    & 1\\
    \bottomrule
  \end{tabular}
\end{table}

\begin{table}[H]
  \caption{BiLSTM+Prot2Vec}
  \label{BiLSTMProt2Vec-hyperparams}
  \centering
  \scriptsize
  \begin{tabular}{lccc}
    \toprule
    \multirow{2}*{\textbf{Hyperparameter}}  & \textbf{Range}  & \multicolumn{2}{c}{\textbf{Final value}} \\
    & (Uniformly random choice) & \textbf{C-terminal} & \textbf{N-terminal}\\
    \midrule
    Epochs trained &    $\leq$ 60 &    60 &    60\\
    Learning rate &    $\{8\times10^{-5}, ,10^{-4}, 3\times10^{-4}, 5\times10^{-4}\}$    & $3\times10^{-4}$ &    $3\times10^{-4}$\\
    Dropout rate &    $\{0.45,0.46,\ldots,0.51,0.52\}$    & 0.5 &    0.5\\
    Linear layer size    & $[120,180)$ &    145    & 139\\
    LSTM size     & $[480,531)$ &    480    & 531\\
    \bottomrule
  \end{tabular}
\end{table}

\begin{table}[H]
  \caption{CNN}
  \label{CNNAttention-hyperparams}
  \centering
  \scriptsize
  \begin{tabular}{lccc}
    \toprule
    \multirow{2}*{\textbf{Hyperparameter}}  & \textbf{Range}  & \multicolumn{2}{c}{\textbf{Final value}} \\
    & (Uniformly random choice) & \textbf{C-terminal} & \textbf{N-terminal}\\
    \midrule
    Epochs trained &    $\leq$ 60 &    60 &    60\\
    Learning Rate &    $\{8\times10^{-5}, 10^{-4}, 3\times10^{-4}, 5\times10^{-4}\}$    & $3\times10^{-4}$ &    $3\times10^{-4}$\\
    Dropout Rate &    $\{0,0.02,\ldots,0.08,0.1\}$    & 0.04 &    0.08\\
    Linear layer size 1    & $[64, 101)$ &    79    & 89\\
    Linear layer size 2    & $[15, 33)$ &    24    & 15\\
    Attention heads 1    & $\{1, 2, 3, 4, 5, 6\}$ &    3    & 4\\
    Attention heads 2    & $\{1, 2, 3, 4, 5, 6\}$ &    2    & 3\\
    Filter size 1 & - & 1 & 1\\
    Number filters 1    &$[220, 301)$    &220    &249\\
    Number filters 2    &$[220, 301)$    &262    &229\\
    Filter size 2a    &$\{1, 3, 5, 7\}$    &3    &1\\
    Filter size 2b    &$\{15, 17, 19, 21, 23, 25\}$    &17    &15\\
    Filter size 3c    &$\{13, 15, 17, 19, 21, 23\}$    &13    &13\\
    Number filters 3    &$[350, 431)$    &398    &400\\
    Filter size 3a    &$\{11, 13, 15, 17, 19\}$    &11    &13\\
    Filter size 3b    &$\{13, 15, 17, 19, 21, 23\}$    &15    &21\\
    Filter size 3c    &$\{11, 13, 15, 17, 19\}$    &19    &15\\
    \bottomrule
  \end{tabular}
\end{table}

\begin{table}[H]
  \caption{MLP}
  \label{MLP-hyperparams}
  \centering
  \scriptsize
  \begin{tabular}{lccc}
    \toprule
    \multirow{2}*{\textbf{Hyperparameter}}  & \textbf{Range}  & \multicolumn{2}{c}{\textbf{Final value}} \\
    & (Uniformly random choice) & \textbf{C-terminal} & \textbf{N-terminal}\\
    \midrule
    Epochs trained &    $\leq$ 60 &    30 &    30\\
    Learning rate &    $\{10^{-4}, 5\times10^{-4},  8\times10^{-4}, 10^{-3}\}$    & $10^{-3}$ &    $10^{-3}$\\
    Dropout rate &    $\{0.1,0.12,\ldots,0.24,0.26\}$    & 0.24 &    0.24\\
    Linear layer size    & $[120, 201)$ &    144    & 167\\
    \bottomrule
  \end{tabular}
\end{table}

\begin{table}[H]
  \caption{BiLSTM+ESM2}
  \label{ESM2EmbedBiLSTM-hyperparams}
  \centering
  \scriptsize
  \begin{tabular}{lcc}
    \toprule
    \multirow{2}*{\textbf{Hyperparameter}} & \multicolumn{2}{c}{\textbf{Final value}} \\
    & \textbf{C-terminal} & \textbf{N-terminal}\\
    \midrule
    Epochs trained &    10 &    10\\
    Learning rate     & $3\times10^{-4}$ &    $3\times10^{-4}$\\
    Dropout rate &    0.5 &    0.5\\
    Linear layer size &    128    & 128\\
    LSTM size & 512 & 512\\
    \bottomrule
  \end{tabular}
\end{table}

\begin{table}[H]
  \caption{ESM2}
  \label{ESM2Finetune-hyperparams}
  \centering
  \scriptsize
  \begin{tabular}{lcc}
    \toprule
    \multirow{2}*{\textbf{Hyperparameter}} & \multicolumn{2}{c}{\textbf{Final value}} \\
    & \textbf{C-terminal} & \textbf{N-terminal}\\
    \midrule
    Epochs trained &    3 &    3\\
    Learning rate     & $2\times10^{-5}$ &    $2\times10^{-5}$\\
    Dropout rate &    0.5 &    0.5\\
    \bottomrule
  \end{tabular}
\end{table}

\begin{table}[H]
  \caption{BiLSTM+T5}
  \label{T5BiLSTM-hyperparams}
  \centering
  \scriptsize
  \begin{tabular}{lcc}
    \toprule
    \multirow{2}*{\textbf{Hyperparameter}} & \multicolumn{2}{c}{\textbf{Final value}} \\
    & \textbf{C-terminal} & \textbf{N-terminal}\\
    \midrule
    Epochs trained &    10 &    10\\
    Learning rate     & $3\times10^{-4}$ &    $3\times10^{-4}$\\
    Dropout rate &    0.5 &    0.5\\
    Linear layer size &    128    & 128\\
    LSTM size & 512 & 512\\
    \bottomrule
  \end{tabular}
\end{table}

\begin{table}[H]
  \caption{BiLSTM+BBPE1, BiLSTM+BBPE50, BiLSTM+WP50}
  \label{tokenizerBiLSTM-hyperparams}
  \centering
  \scriptsize
  \begin{tabular}{lcc}
    \toprule
    \multirow{2}*{\textbf{Hyperparameter}} & \multicolumn{2}{c}{\textbf{Final value}} \\
    & \textbf{C-terminal} & \textbf{N-terminal}\\
    \midrule
    Epochs trained &    15 &    15\\
    Learning rate     & $10^{-4}$ &    $10^{-4}$\\
    Dropout rate &    0.5 &    0.5\\
    Embedding dimension & 150 & 150\\
    Linear layer size &    128    & 128\\
    LSTM size & 512 & 512\\
    \bottomrule
  \end{tabular}
\end{table}

\begin{table}[H]
  \caption{BiLSTM+FwBw}
  \label{FwBwBiLSTM-hyperparams}
  \centering
  \scriptsize
  \begin{tabular}{lcc}
    \toprule
    \multirow{2}*{\textbf{Hyperparameter}} & \multicolumn{2}{c}{\textbf{Final value}} \\
    & \textbf{C-terminal} & \textbf{N-terminal}\\
    \midrule
    Epochs trained &    15 &    15\\
    Learning rate     & $10^{-4}$ &    $10^{-4}$\\
    Dropout rate &    0.5 &    0.5\\
    Linear layer size 1 &    128    & 128\\
    % Embedding dimension & 100+100 & 100+100\\
    LSTM size 1 & 128 & 128\\
    LSTM size 2 & 512 & 512\\
    Sequence encoding embedding dimension & 100 & 100\\
    Sequence encoding BiLSTM size & 200 &200\\
    \bottomrule
  \end{tabular}
\end{table}

\begin{table}[H]
  \caption{Co-Teaching, Co-Teaching+, JoCoR}
  \label{ccj-hyperparams}
  \centering
  \scriptsize
  \begin{tabular}{lccc}
    \toprule
    \multirow{2}*{\textbf{Hyperparameter}} & \multicolumn{3}{c}{\textbf{Final value}} \\
    & \textbf{Co-teaching} & \textbf{Co-teaching+} & \textbf{JoCoR}\\
    \midrule
    Number scale-up epochs    &10    &10    &10\\
    Noise rate    &0.2    &0.2    &0.2\\
    Forget rate    &0.2    &0.2    &0.1\\
    Exponent    &1    &1    &1\\
    \bottomrule
  \end{tabular}
\end{table}

\begin{table}[H]
  \caption{DivideMix}
  \label{dividemix-hyperparams}
  \centering
  \scriptsize
  \begin{tabular}{lc}
    \toprule
    \textbf{Hyperparameter}  & \textbf{Final value}  \\
    \midrule
    Number warm-up epochs    &1\\
    $\alpha$    &0.5\\
    $\lambda_{u}$    &0\\
    Probability threshold    &0.5\\
    Temperature    &0.5\\
    Number scale-up epochs    &5\\
    \bottomrule
  \end{tabular}
\end{table}

\begin{table}[H]
  \caption{Noise adaptation layer}
  \label{noiselayer-hyperparams}
  \centering
  \scriptsize
  \begin{tabular}{lc}
    \toprule
    \textbf{Hyperparameter}  & \textbf{Final value}  \\
    \midrule
    Number warm-up epochs    &1\\
    $\beta$    &0.8\\
    \bottomrule
  \end{tabular}
\end{table}

\subsection{Results without denoising methods}
\label{apx:resultsnodenoise}

\begin{table}[H]
  \caption{Model performances on C- and N-terminals}
  \label{c-nodenoise}
  \centering
  \scriptsize
  \begin{tabular}{crcccc}
    \toprule
    & & \multicolumn{2}{c}{\textbf{C-terminal}} & \multicolumn{2}{c}{\textbf{N-terminal}} \\
    \textbf{Priority} & \textbf{Models} & \textbf{AUC} & \textbf{ACC} & \textbf{AUC} & \textbf{ACC} \\
    \cmidrule(r){1-2}\cmidrule(){3-4}\cmidrule(l){5-6}
    \multirow{5}*{1}
    & BiLSTM  & 88.55 $\pm$ 0.12 & 79.50 $\pm$ 0.11 &79.50 $\pm$ 0.11    &83.51 $\pm$ 0.11    \\
    & BiLSTM+Attention   & 88.28 $\pm$ 0.08 & 79.24 $\pm$ 0.11 & 79.24 $\pm$ 0.11    &83.36 $\pm$ 0.13    \\
    & BiLSTM+Prot2Vec   & 87.99 $\pm$ 0.14 & 79.10 $\pm$ 0.11 & 79.10 $\pm$ 0.11    &83.22 $\pm$ 0.13    \\
    & CNN   & 86.66 $\pm$ 0.17 & 77.30 $\pm$ 0.82 & 77.30 $\pm$ 0.82    &82.89 $\pm$ 0.22    \\
    & MLP   & 88.17 $\pm$ 0.11 & 79.08 $\pm$ 0.11 & 79.08 $\pm$ 0.11    &83.33 $\pm$ 0.12    \\
    \cmidrule(r){1-2}\cmidrule(){3-4}\cmidrule(l){5-6}
    \multirow{3}*{2} 
    & BiLSTM+ESM2  & 88.34 $\pm$ 0.05 & 79.24 $\pm$ 0.10 & 79.24 $\pm$ 0.10    &83.35 $\pm$ 0.09    \\
    & ESM2   & 88.32 $\pm$ 0.16 & 78.91 $\pm$ 0.18 & 78.91 $\pm$ 0.18    &82.63 $\pm$ 0.64    \\
    & BiLSTM+T5  & 88.32 $\pm$ 0.05 & 79.48 $\pm$ 0.11 & 79.48 $\pm$ 0.11    &83.45 $\pm$ 0.08    \\
    \cmidrule(r){1-2}\cmidrule(){3-4}\cmidrule(l){5-6}
    \multirow{4}*{3}
    & BiLSTM+BBPE1   & 85.25 \phantom{$\pm$ 0.00} & 76.56 \phantom{$\pm$ 0.00} & 76.56    \phantom{$\pm$ 0.00}&82.88    \phantom{$\pm$ 0.00} \\
    & BiLSTM+BBPE50   & 68.92 \phantom{$\pm$ 0.00}& 68.67 \phantom{$\pm$ 0.00} & 68.67    \phantom{$\pm$ 0.00}&82.03    \phantom{$\pm$ 0.00} \\
    & BiLSTM+WP50  & 73.46 \phantom{$\pm$ 0.00}& 69.28 \phantom{$\pm$ 0.00} & 69.28    \phantom{$\pm$ 0.00}&82.08    \phantom{$\pm$ 0.00}\\
    & BiLSTM+FwBw   & 87.59 \phantom{$\pm$ 0.00}& 78.71 \phantom{$\pm$ 0.00} & 78.71    \phantom{$\pm$ 0.00}&83.15    \phantom{$\pm$ 0.00}\\
    \bottomrule
  \end{tabular}
\end{table}

\subsection{Results with denoising methods}
\label{apx:resultswithdenoise}

\begin{table}[H]
  \caption{BiLSTM with denoising on C-terminal}
  \label{bilstm-denoise-c}
  \centering
  \scriptsize
  \begin{tabular}{lcccccc}
    \toprule
    \textbf{Denoising methods}  & \textbf{AUC} & \textbf{ACC} & \textbf{Time (s/epoch)} & \textbf{Epochs} & \textbf{Parameters} & \textbf{Trainable parameters} \\
    \midrule
    Co-Teaching    & 87.50    &86.64    &~41 &    15    & \num{4655984}    &\num{4655984} \\
    Co-Teaching+ &    87.50    & 86.64    &~41 &    15    &\num{4655984}    &\num{4655984} \\
    JoCoR    & 84.53    & 85.49    &~41 &    15    & \num{4655984}    & \num{4655984} \\
    DivideMix    & 86.25    & 84.02    & ~210    & 15    & \num{4656149}    &\num{4656149} \\
    Noise Adaptation Layer &    88.49 &    87.02 &    ~22 &    15    & \num{4656149}    & \num{4656149} \\
    \bottomrule
  \end{tabular}
\end{table}

\begin{table}[H]
  \caption{BiLSTM with denoising on N-terminal}
  \label{bilstm-denoise-n}
  \centering
  \scriptsize
  \begin{tabular}{lcccccc}
    \toprule
    \textbf{Denoising methods}  & \textbf{AUC} & \textbf{ACC} & \textbf{Time (s/epoch)} & \textbf{Epochs} & \textbf{Parameters} & \textbf{Trainable parameters} \\
    \midrule
    Co-Teaching    & 78.28    &83.21    &~44 &    15    & \num{5096135}    & \num{5096135} \\
    Co-Teaching+ &    78.37    &83.17    &~44 &    15    &\num{5096135}    &\num{5096135} \\
    JoCoR    & 74.26    &82.12    &~44 &    15    &\num{5096135}    &\num{5096135} \\
    DivideMix    & 77.08    &81.52    &~210 &    15    &\num{5096315}    &\num{5096315} \\
    Noise Adaptation Layer & 79.48    &83.43    &~23 &    15    &\num{5096315}    & \num{5096315} \\
    \bottomrule
  \end{tabular}
\end{table}

\begin{table}[H]
  \caption{BiLSTM+Attention with denoising on C-terminal}
  \label{bilstmattention-denoise-c}
  \centering
  \scriptsize
  \begin{tabular}{lcccccc}
    \toprule
    \textbf{Denoising methods}  & \textbf{AUC} & \textbf{ACC} & \textbf{Time (s/epoch)} & \textbf{Epochs} & \textbf{Parameters} & \textbf{Trainable parameters} \\
    \midrule
    Co-Teaching    &86.82    &86.20    &~40 &    20    &\num{1632391}    &\num{1632391}\\
    Co-Teaching+    &86.91    &86.16    &~40 &    20    &\num{1632391}    &\num{1632391}\\
    JoCoR    &81.70    &85.04    &~38 &    20    &\num{1632391}    &\num{1632391}\\
    Noise Adaptation Layer    &88.23    &86.90    &~18 &    20    &\num{1632539}    &\num{1632539}\\
    \bottomrule
  \end{tabular}
\end{table}

\begin{table}[H]
  \caption{BiLSTM+Attention with denoising on N-terminal}
  \label{bilstmattention-denoise-n}
  \centering
  \scriptsize
  \begin{tabular}{lcccccc}
    \toprule
    \textbf{Denoising methods}  & \textbf{AUC} & \textbf{ACC} & \textbf{Time (s/epoch)} & \textbf{Epochs} & \textbf{Parameters} & \textbf{Trainable parameters} \\
    \midrule
    Co-Teaching    &77.48    &83.13    &~43 &    20    &\num{1718233}    &\num{1718233}\\
    Co-Teaching+    &78.09    &83.13    &~42 &    20    &\num{1718233}    &\num{1718233}\\
    JoCoR    &68.86    &82.40    &~41 &    20    &\num{1718233}    &\num{1718233}\\
    Noise Adaptation Layer    &79.21    &83.37    &~22 &    20    &\num{1718384}    &\num{1718384}\\
    \bottomrule
  \end{tabular}
\end{table}

\begin{table}[H]
  \caption{BiLSTM+Prot2Vec with denoising on C-terminal}
  \label{bilstmprot2vec-denoise-c}
  \centering
  \scriptsize
  \begin{tabular}{lcccccc}
    \toprule
    \textbf{Denoising methods}  & \textbf{AUC} & \textbf{ACC} & \textbf{Time (s/epoch)} & \textbf{Epochs} & \textbf{Parameters} & \textbf{Trainable parameters} \\
    \midrule
    Co-Teaching    &85.47    &85.78    &~32 &    15    &\num{16009371}    &\num{5830371}\\
    Co-Teaching+    &85.47    &85.78    &~31 &    15    &\num{16009371}    &\num{5830371}\\
    JoCoR    &75.80    &82.20    &~32 &    15    &\num{16009371}    &\num{5830371}\\
    Noise Adaptation Layer    &87.93    &86.60    &~15 &    15    &\num{16009517}    &\num{5830517}\\
    \bottomrule
  \end{tabular}
\end{table}

\begin{table}[H]
  \caption{BiLSTM+Prot2Vec with denoising on N-terminal}
  \label{bilstmprot2vec-denoise-n}
  \centering
  \scriptsize
  \begin{tabular}{lcccccc}
    \toprule
    \textbf{Denoising methods}  & \textbf{AUC} & \textbf{ACC} & \textbf{Time (s/epoch)} & \textbf{Epochs} & \textbf{Parameters} & \textbf{Trainable parameters} \\
    \midrule
    Co-Teaching    &76.64    &83.00    &~40 &    15    &\num{16772049}    &\num{6593049}\\
    Co-Teaching+    &77.44    &82.90    &~40 &    15    &\num{16772049}    &\num{6593049}\\
    JoCoR    &61.62    &81.91    &~40 &    15    &\num{16772049}    &\num{6593049}\\
    Noise Adaptation Layer    &78.96    &83.15    &~19 &    15    &\num{16772189}    &\num{6593189}\\
    \bottomrule
  \end{tabular}
\end{table}

\begin{table}[H]
  \caption{CNN with denoising on C-terminal}
  \label{cnn-denoise-c}
  \centering
  \scriptsize
  \begin{tabular}{lcccccc}
    \toprule
    \textbf{Denoising methods}  & \textbf{AUC} & \textbf{ACC} & \textbf{Time (s/epoch)} & \textbf{Epochs} & \textbf{Parameters} & \textbf{Trainable parameters} \\
    \midrule
    Co-Teaching    &71.51    &82.35    &~102 &    60    &\num{16084198}    &\num{16084198}\\
    Co-Teaching+    &76.18    &82.91    &~102 &    60    &\num{16084198}    &\num{16084198}\\
    JoCoR    &71.10    &82.92    &~102 &    60    &\num{16084198}    &\num{16084198}\\
    Noise Adaptation Layer    &85.91    &85.50    &~49 &    60    &\num{16084223}    & \num{16084223}\\
    \bottomrule
  \end{tabular}
\end{table}

\begin{table}[H]
  \caption{CNN with denoising on N-terminal}
  \label{cnn-denoise-n}
  \centering
  \scriptsize
  \begin{tabular}{lcccccc}
    \toprule
    \textbf{Denoising methods}  & \textbf{AUC} & \textbf{ACC} & \textbf{Time (s/epoch)} & \textbf{Epochs} & \textbf{Parameters} & \textbf{Trainable parameters} \\
    \midrule
    Co-Teaching    &60.71    &81.88    &~102&    60    &\num{15237057} &    \num{15237057}\\
    Co-Teaching+ &74.15    &82.05    &~102&    60    &\num{15237057}    &\num{15237057}\\
    JoCoR &60.81    &81.88    &~102 &    60    &\num{15237057}    &\num{15237057}\\
    Noise Adaptation Layer    &76.68    &82.78    &~47 &    60    &\num{15237073}    &\num{15237073}\\
    \bottomrule
  \end{tabular}
\end{table}

\begin{table}[H]
  \caption{MLP with denoising on C-terminal}
  \label{mlp-denoise-c}
  \centering
  \scriptsize
  \begin{tabular}{lcccccc}
    \toprule
    \textbf{Denoising methods}  & \textbf{AUC} & \textbf{ACC} & \textbf{Time (s/epoch)} & \textbf{Epochs} & \textbf{Parameters} & \textbf{Trainable parameters} \\
    \midrule
    Co-Teaching    &87.14    &86.45    &~6 &    30    &\num{30529}    &\num{30529}\\
    Co-Teaching+    &87.16    &85.85    &~6 &    30    &\num{30529}    &\num{30529}\\
    JoCoR    &86.11    &85.56    &~5 &    30    &\num{30529}    &\num{30529}\\
    Noise Adaptation Layer    &88.07    &86.73    &~3 &    30    &\num{30674}    &\num{30674}\\
    \bottomrule
  \end{tabular}
\end{table}

\begin{table}[H]
  \caption{MLP with denoising on N-terminal}
  \label{mlp-denoise-n}
  \centering
  \scriptsize
  \begin{tabular}{lcccccc}
    \toprule
    \textbf{Denoising methods}  & \textbf{AUC} & \textbf{ACC} & \textbf{Time (s/epoch)} & \textbf{Epochs} & \textbf{Parameters} & \textbf{Trainable parameters} \\
    \midrule
    Co-Teaching    &77.89    &83.21    &~6 &    30    &\num{35405}    &\num{35405}\\
    Co-Teaching+    &78.34    &83.11    &~5 &    30    &\num{35405}    &\num{35405}\\
    JoCoR    &76.31    &82.30    &~5 &    30    &\num{35405}    &\num{35405}\\
    Noise Adaptation Layer    &78.82    &83.32    &~4 &    30    &\num{35573}    &\num{35573}\\
    \bottomrule
  \end{tabular}
\end{table}

\begin{table}[H]
  \caption{BiLSTM+ESM2 with denoising on C-terminal}
  \label{ESM2EmbedBiLSTM-denoise-c}
  \centering
  \scriptsize
  \begin{tabular}{lcccccc}
    \toprule
    \textbf{Denoising methods}  & \textbf{AUC} & \textbf{ACC} & \textbf{Time (s/epoch)} & \textbf{Epochs} & \textbf{Parameters} & \textbf{Trainable parameters} \\
    \midrule
    Co-Teaching    &87.19    &86.52    &~660    &10    &\num{152998267}    &\num{4858113}\\
    Co-Teaching+    &87.19    &86.52    &~660 &    10    &\num{152998267}    &\num{4858113}\\
    JoCoR    &84.52    &84.93    &~660 &    10    &\num{152998267}    &\num{4858113}\\
    Noise Adaptation Layer    &88.32    &86.88    &~330    &10    &\num{152998396}    &\num{4858242}\\
    \bottomrule
  \end{tabular}
\end{table}

\begin{table}[H]
  \caption{BiLSTM+ESM2 with denoising on N-terminal}
  \label{ESM2EmbedBiLSTM-denoise-n}
  \centering
  \scriptsize
  \begin{tabular}{lcccccc}
    \toprule
    \textbf{Denoising methods}  & \textbf{AUC} & \textbf{ACC} & \textbf{Time (s/epoch)} & \textbf{Epochs} & \textbf{Parameters} & \textbf{Trainable parameters} \\
    \midrule
    Co-Teaching    &77.76    &83.25    &~660 &    10    &\num{152998267}    &\num{4858113}\\
    Co-Teaching+    &78.03    &83.15    &~660 &    10    &\num{152998267}    &\num{4858113}\\
    JoCoR    &73.71    &81.96    &~660 &    10    &\num{152998267}    &\num{4858113}\\
    Noise Adaptation Layer    &79.29    &83.37    &~360 &    10    &\num{152998396}    &\num{4858242}\\
    \bottomrule
  \end{tabular}
\end{table}

\begin{table}[H]
  \caption{ESM2 with denoising on C-terminal}
  \label{ESM2FineTuned-denoise-c}
  \centering
  \scriptsize
  \begin{tabular}{lcccccc}
    \toprule
    \textbf{Denoising methods}  & \textbf{AUC} & \textbf{ACC} & \textbf{Time (s/epoch)} & \textbf{Epochs} & \textbf{Parameters} & \textbf{Trainable parameters} \\
    \midrule
    Co-Teaching    &87.93    &85.56    &\num{2160} &    3    &\num{148140188}    &\num{148140188}\\
    Co-Teaching+    &87.93    &85.56    &\num{2100} &    3    &\num{148140188}    &\num{148140188}\\
    JoCoR    &87.31    &86.25    &\num{2100} &    3    &\num{148140188}    &\num{148140188}\\
    Noise Adaptation Layer    &87.97    &86.58    &~960 &    3    &\num{148140222}    &\num{148140222}\\
    \bottomrule
  \end{tabular}
\end{table}

\begin{table}[H]
  \caption{ESM2 with denoising on N-terminal}
  \label{ESM2FineTuned-denoise-n}
  \centering
  \scriptsize
  \begin{tabular}{lcccccc}
    \toprule
    \textbf{Denoising methods}  & \textbf{AUC} & \textbf{ACC} & \textbf{Time (s/epoch)} & \textbf{Epochs} & \textbf{Parameters} & \textbf{Trainable parameters} \\
    \midrule
    Co-Teaching    &78.53    &80.54    &\num{2220} &    3    &\num{148140188}    &\num{148140188}\\
    Co-Teaching+    &78.53    &80.54    &\num{2220} &    3 &    \num{148140188}    &\num{148140188}\\
    JoCoR    &77.83    &83.02    &\num{2160} &    3    &\num{148140188}    &\num{148140188}\\
    Noise Adaptation Layer    &77.31    &82.13    &~930 &    3    &\num{148140222}    &\num{148140222}\\
    \bottomrule
  \end{tabular}
\end{table}

\begin{table}[H]
  \caption{BiLSTM+T5 with denoising on C-terminal}
  \label{T5BiLSTM-denoise-c}
  \centering
  \scriptsize
  \begin{tabular}{lcccccc}
    \toprule
    \textbf{Denoising methods}  & \textbf{AUC} & \textbf{ACC} & \textbf{Time (s/epoch)} & \textbf{Epochs} & \textbf{Parameters} & \textbf{Trainable parameters} \\
    \midrule
    Co-Teaching    &87.11    &86.40    &\num{1500} &    10    &\num{1214572801}    &\num{6430977}\\
    Co-Teaching+    &87.11    &86.40    &\num{1500} &    10    &\num{1214572801}    &\num{6430977}\\
    JoCoR    &83.48    &83.88    &\num{1500} &    10    &\num{1214572801}    &\num{6430977}\\
    Noise Adaptation Layer    &88.36    &86.85    &~720 &     10    &\num{1214572930}    &\num{6431106}\\
    \bottomrule
  \end{tabular}
\end{table}

\begin{table}[H]
  \caption{BiLSTM+T5 with denoising on N-terminal}
  \label{T5BiLSTM-denoise-n}
  \centering
  \scriptsize
  \begin{tabular}{lcccccc}
    \toprule
    \textbf{Denoising methods}  & \textbf{AUC} & \textbf{ACC} & \textbf{Time (s/epoch)} & \textbf{Epochs} & \textbf{Parameters} & \textbf{Trainable parameters} \\
    \midrule
    Co-Teaching    &78.30    &83.22    &\num{1560} &    10    &\num{1214572801}    &\num{6430977}\\
    Co-Teaching+    &78.32    &83.26    &\num{1560} &    10    &\num{1214572801}    &\num{6430977}\\
    JoCoR    &71.76    &82.28    &\num{1500} &    10    &\num{1214572801}    &\num{6430977}\\
    Noise Adaptation Layer    &79.54    &83.48    &~780 &    10    &\num{1214572930}    &\num{6431106}\\
    \bottomrule
  \end{tabular}
\end{table}

\begin{table}[H]
  \caption{BiLSTM+BBPE1 with denoising on C-terminal}
  \label{BiLSTMbbpe1k-denoise-c}
  \centering
  \scriptsize
  \begin{tabular}{lcccccc}
    \toprule
    \textbf{Denoising methods}  & \textbf{AUC} & \textbf{ACC} & \textbf{Time (s/epoch)} & \textbf{Epochs} & \textbf{Parameters} & \textbf{Trainable parameters} \\
    \midrule
    Co-Teaching    &76.55    &83.84    &~33 &    15    &\num{5319409}    &\num{5319409}\\
    Co-Teaching+    &80.31    &83.44    &~34 &    15    &\num{5319409}    &\num{5319409}\\
    JoCoR    &58.81    &82.20    &~33 &    15    &\num{5319409}    &\num{5319409}\\
    Noise Adaptation Layer    &85.15    &85.45    &~17 &    15    &\num{5319538}    &\num{5319538}\\
    \bottomrule
  \end{tabular}
\end{table}

\begin{table}[H]
  \caption{BiLSTM+BBPE1 with denoising on N-terminal}
  \label{BiLSTMbbpe1k-denoise-n}
  \centering
  \scriptsize
  \begin{tabular}{lcccccc}
    \toprule
    \textbf{Denoising methods}  & \textbf{AUC} & \textbf{ACC} & \textbf{Time (s/epoch)} & \textbf{Epochs} & \textbf{Parameters} & \textbf{Trainable parameters} \\
    \midrule
    Co-Teaching    &69.12    &82.19    &~34 &    15    &\num{5319409}    &\num{5319409}\\
    Co-Teaching+    &74.65    &82.37    &~34 &    15    &\num{5319409}    &\num{5319409}\\
    JoCoR    &58.03    &81.88    &~33 &    15    &\num{5319409}    &\num{5319409}\\
    Noise Adaptation Layer    &76.15    &82.58    &~17 &    15    &\num{5319538}    &\num{5319538}\\
    \bottomrule
  \end{tabular}
\end{table}

\begin{table}[H]
  \caption{BiLSTM+BBPE50 with denoising on C-terminal}
  \label{BiLSTMbbpe50k-denoise-c}
  \centering
  \scriptsize
  \begin{tabular}{lcccccc}
    \toprule
    \textbf{Denoising methods}  & \textbf{AUC} & \textbf{ACC} & \textbf{Time (s/epoch)} & \textbf{Epochs} & \textbf{Parameters} & \textbf{Trainable parameters} \\
    \midrule
    Co-Teaching    &57.62    &82.28    &~30 &    15    &\num{12669409}    &\num{12669409}\\
    Co-Teaching+    &67.40    &82.35    &~30 &    15    &\num{12669409}    &\num{12669409}\\
    JoCoR    &52.09    &82.20    &~30 &    15    &\num{12669409}    &\num{12669409}\\
    Noise Adaptation Layer    &67.42    &82.38    &~16 &    15    &\num{12669538}    &\num{12669538}\\
    \bottomrule
  \end{tabular}
\end{table}

\begin{table}[H]
  \caption{BiLSTM+BBPE50 with denoising on N-terminal}
  \label{BiLSTMbbpe50k-denoise-n}
  \centering
  \scriptsize
  \begin{tabular}{lcccccc}
    \toprule
    \textbf{Denoising methods}  & \textbf{AUC} & \textbf{ACC} & \textbf{Time (s/epoch)} & \textbf{Epochs} & \textbf{Parameters} & \textbf{Trainable parameters} \\
    \midrule
    Co-Teaching    &62.52    &81.88    &~31 &    15    &\num{12669409}    &\num{12669409}\\
    Co-Teaching+    &67.45    &81.98    &~32 &    15    &\num{12669409}    &\num{12669409}\\
    JoCoR    &53.02    &81.88    &~31 &    15    &\num{12669409}    &\num{12669409}\\
    Noise Adaptation Layer    &67.90    &81.95    &~15 &    15    &\num{12669538}    &\num{12669538}\\
    \bottomrule
  \end{tabular}
\end{table}

\begin{table}[H]
  \caption{BiLSTM+WP50 with denoising on C-terminal}
  \label{BiLSTMbwp-denoise-c}
  \centering
  \scriptsize
  \begin{tabular}{lcccccc}
    \toprule
    \textbf{Denoising methods}  & \textbf{AUC} & \textbf{ACC} & \textbf{Time (s/epoch)} & \textbf{Epochs} & \textbf{Parameters} & \textbf{Trainable parameters} \\
    \midrule
    Co-Teaching    &59.31    &82.25    &~40 &    15    &\num{12669409}    &\num{12669409}\\
    Co-Teaching+    &69.37    &82.57    &~40 &    15    &\num{12669409}    &\num{12669409}\\
    JoCoR    &53.45    &82.20    &~39 &    15    &\num{12669409}    &\num{12669409}\\
    Noise Adaptation Layer    &72.80    &82.98    &~20 &    15    &\num{12669538}    &\num{12669538}\\
    \bottomrule
  \end{tabular}
\end{table}

\begin{table}[H]
  \caption{BiLSTM+WP50 with denoising on N-terminal}
  \label{BiLSTMbwp-denoise-n}
  \centering
  \scriptsize
  \begin{tabular}{lcccccc}
    \toprule
    \textbf{Denoising methods}  & \textbf{AUC} & \textbf{ACC} & \textbf{Time (s/epoch)} & \textbf{Epochs} & \textbf{Parameters} & \textbf{Trainable parameters} \\
    \midrule
    Co-Teaching    &62.66    &81.92    &~37 &    15    &\num{12669409}    &\num{12669409}\\
    Co-Teaching+    &68.44    &81.97    &~37 &    15    &\num{12669409}    &\num{12669409}\\
    JoCoR    &56.25    &81.88    &~36 &    15    &\num{12669409}    &\num{12669409}\\
    Noise Adaptation Layer    &68.91    &81.99    &~20 &    15    &\num{12669538}    &\num{12669538}\\
    \bottomrule
  \end{tabular}
\end{table}

\begin{table}[H]
  \caption{BiLSTM+FwBw with denoising on C-terminal}
  \label{FwBwBiLSTM-denoise-c}
  \centering
  \scriptsize
  \begin{tabular}{lcccccc}
    \toprule
    \textbf{Denoising methods}  & \textbf{AUC} & \textbf{ACC} & \textbf{Time (s/epoch)} & \textbf{Epochs} & \textbf{Parameters} & \textbf{Trainable parameters} \\
    \midrule
    Co-Teaching    &85.84    &84.92    &~270 &    15    &\num{4315369}    &\num{4315369}\\
    Co-Teaching+    &85.84    &84.92    &~270 &    15    &\num{4315369}    &\num{4315369}\\
    JoCoR    &61.26    &82.20    &~270 &    15    &\num{4315369}    &\num{4315369}\\
    Noise Adaptation Layer    &87.75    &86.09    &~120 &    15    &\num{4315498}    &\num{4315498}\\
    \bottomrule
  \end{tabular}
\end{table}

\begin{table}[H]
  \caption{BiLSTM+FwBw with denoising on N-terminal}
  \label{FwBwBiLSTM-denoise-n}
  \centering
  \scriptsize
  \begin{tabular}{lcccccc}
    \toprule
    \textbf{Denoising methods}  & \textbf{AUC} & \textbf{ACC} & \textbf{Time (s/epoch)} & \textbf{Epochs} & \textbf{Parameters} & \textbf{Trainable parameters} \\
    \midrule
    Co-Teaching    &76.86    &82.45    &~270 &    15    &\num{4315369}    &\num{4315369}\\
    Co-Teaching+    &77.02    &82.88    &~270 &    15    &\num{4315369}    &\num{4315369}\\
    JoCoR    &62.00    &81.88    &~270 &    15    &\num{4315369}    &\num{4315369}\\
    Noise Adaptation Layer    &78.64    &83.09    &~120 &15    &\num{4315498}    &\num{4315498}\\
    \bottomrule
  \end{tabular}
\end{table}

\end{document}